\documentclass[sn-aps]{sn-jnl}
\jyear{2022}
\usepackage{caption}
\usepackage{subcaption}
\usepackage{adjustbox}

\begin{document}
\title[ ]{Measurements of the Cherenkov effect in direct detection of charged particles with SiPMs}

\author*[1]{\fnm{F.} \sur{Carnesecchi}}\email{francesca.carnesecchi@cern.ch}

\author[2,3] {\footnote[3]{Now at ICSC, World Laboratory, c\textbackslash o CERN Geneva }   \fnm{B.} \sur{Sabiu}}
\author[2,3]{\fnm{S.} \sur{Strazzi}}
\author[3]{\footnote[4]{Now at Deutsches Elektronen-Synchrotron DESY, Hamburg }  \fnm{G.} \sur{Vignola}}
\author[2,3]{\fnm{N.} \sur{Agrawal}}
\author[2,3]{\fnm{A.} \sur{Alici}}
\author[3]{\fnm{P.} \sur{Antonioli}}
\author[2,3]{\fnm{S.} \sur{Arcelli}}
\author[2,3]{\fnm{F.} \sur{Bellini}}
\author[3]{\fnm{D.} \sur{Cavazza}}
\author[2,3]{\fnm{L.} \sur{Cifarelli}}
\author[2,3]{\fnm{M.} \sur{Colocci}}
\author[4]{\fnm{S.} \sur{Durando}}
\author[2,3]{\fnm{F.} \sur{Ercolessi}}
\author[3]{\fnm{D.} \sur{Falchieri}}
\author[5]{\fnm{A.} \sur{Ficorella}}
\author[2,3]{\fnm{C.} \sur{Fraticelli}}
\author[6,3]{\fnm{M.} \sur{Garbini}}
\author[2,3]{\fnm{M.} \sur{Giacalone}}
\author[5]{\fnm{A.} \sur{Gola}}
\author[3]{\fnm{D.} \sur{Hatzifotiadou}}
\author[2,3]{\fnm{N.} \sur{Jacazio}}
\author[3]{\fnm{A.} \sur{Margotti}}
\author[2,3]{\fnm{G.} \sur{Malfattore}}
\author[3]{\fnm{R.} \sur{Nania}}
\author[3]{\fnm{F.} \sur{Noferini}}
\author[5]{\fnm{G.} \sur{Paternoster}}
\author[3]{\fnm{O.} \sur{Pinazza}}
\author[3]{\fnm{R.} \sur{Preghenella}}
\author[3]{\fnm{R.} \sur{Rath}}
\author[7]{\fnm{R.} \sur{Ricci}}
\author[3]{\fnm{L.} \sur{Rignanese}}
\author[2,3]{\fnm{G.} \sur{Romanenko}}
\author[2,3]{\fnm{N.} \sur{Rubini}}

\author[3]{\fnm{E.} \sur{Scapparone}}
\author[2,3]{\fnm{G.} \sur{Scioli}}
\author[2,3]{\fnm{A.} \sur{Zichichi}}

\affil[1]{\orgname{CERN}, \orgaddress{\street{Esplanade des Particules 1}, \city{Geneva}, \postcode{1211 Geneva 23},  \country{Switzerland}}}

\affil[2]{\orgdiv{Dipartimento di Fisica e Astronomia "A. Righi"}, \orgname{University of Bologna}, \orgaddress{\street{viale Carlo Berti Pichat 6/2}, \city{Bologna}, \postcode{40127}, \country{Italy}}}

\affil[3]{\orgdiv{Sezione di Bologna}, \orgname{Istituto Nazionale di Fisica Nucleare}, \orgaddress{\street{viale Carlo Berti Pichat 6/2}, \city{Bologna}, \postcode{40127}, \country{Italy}}}

\affil[4]{\orgdiv{Dipartimento di elettronica e telecomunicazioni}, \orgname{Politecnico di
Torino}, \orgaddress{\street{Corso Duca degli Abruzzi, 24}, \city{Torino}, \postcode{10129}, \country{Italy}}}

\affil[5]{\orgname{Fondazione Bruno Kessler}, \orgaddress{\street{Via Sommarive, 18}}, \orgaddress{\city{Povo}, \postcode{38123}, \country{Italy}}}

\affil[6]{ \orgname{Museo Storico della Fisica e Centro Studi Enrico Fermi}, \orgaddress{\street{Via Panisperna 89 A}, \city{Roma}, \postcode{10129}, \country{Italy}}}

\affil[7]{\orgdiv{Dipartmento di Fisica }, \orgname{University of Salerno}, \orgaddress{\street{Via Giovanni Paolo II, 132}, \city{Salerno}, \postcode{84084}, \country{Italy}}}


\abstract{In this paper, different Silicon PhotoMultiplier (SiPM) sensors have been tested with charged particles to characterize the Cherenkov light produced in the sensor protection layer. 
A careful position scan of the SiPM response has been performed with different prototypes, confirming the large number of firing cells and proving almost full efficiency, with the SiPM filling factor essentially negligible. This study also allowed us to study the time resolution of such devices as a function of the number of firing cells, reaching values below 20 ps.
These measurements provide significant insight into the capabilities of SiPM sensors in direct detection of charged particles and their potential for several applications.
}


\maketitle

\section{Introduction}\label{sec1}

In  \cite{SiPM1} it was quantitatively shown that the response of different Silicon PhotoMultipliers (SiPMs) to the passage of a charged particle was characterized by an excess of firing cells (SPADs) with respect to the expectations from CrossTalk (CT). In \cite{SiPM2} it was shown that the effect was not related to internal silicon bulk, but to the presence of a protection layer where, at the passage of the charged particle, the Cherenkov effect produces a significant amount of photons that arrive at the same time on the SiPM surface generating a large signal with many firing SPADs. In both papers time resolution measurements were reported. In particular, reference \cite{SiPM2} showed an improvement with the number of fired SPADs,  reaching values of about 30-40 ps for $\geq$ 4 SPADs.

In this work, results of a new beam test are reported where a position scan was performed on SiPMs with different protection layers, aiming to compare them both in terms of SPADs distribution and time resolution, well beyond the previous range of measured SPADs.

This work is part of the R\&D studies for the TOF timing layer of the ALICE 3 \cite{ALICE3} experiment at the LHC.

\section{Experimental setup}\label{sec2}

\subsection{Detectors} 
\label{sec:detectors}
For the present study available NUV-HD-RH SiPMs produced by Fondazione Bruno Kessler (FBK) were used  \cite{2020Mazzi}.
These detectors are based on the NUV-HD technology \cite{2019Gola} and have an active area of 1$\times$1 mm$^2$, hexagonal pixel, with an equivalent rectangular pixel pitch of 20 $\mu \text{m}$, 
2444 SPADs,  72$\%$ fill factor and
V$_{bd}$  33.0 $\pm$0.1 V \cite{SiPM1}.  The detectors were produced with standard protection layers; in particular, in this paper three different protection layers of 1.0 and 1.5 mm silicon resin (refraction index 1.5, named SR1 and SR15)  and 1.0 mm epoxy resin (refraction index 1.53, named ER1), thickness with respect to the board hosting the sensor,  have been studied. Since the sensor itself is 550 $\mu$m thick, the two thicknesses correspond to an effective protection layer on top of the sensor of 450  and 950 $\mu$m  respectively for the 1.0 and 1.5 mm.  A fourth prototype was produced without any protection resin (named WR) and was the same used in \cite{SiPM2}. 

\begin{figure}[h!]
        \centering%
        \includegraphics [width=0.8\textwidth]{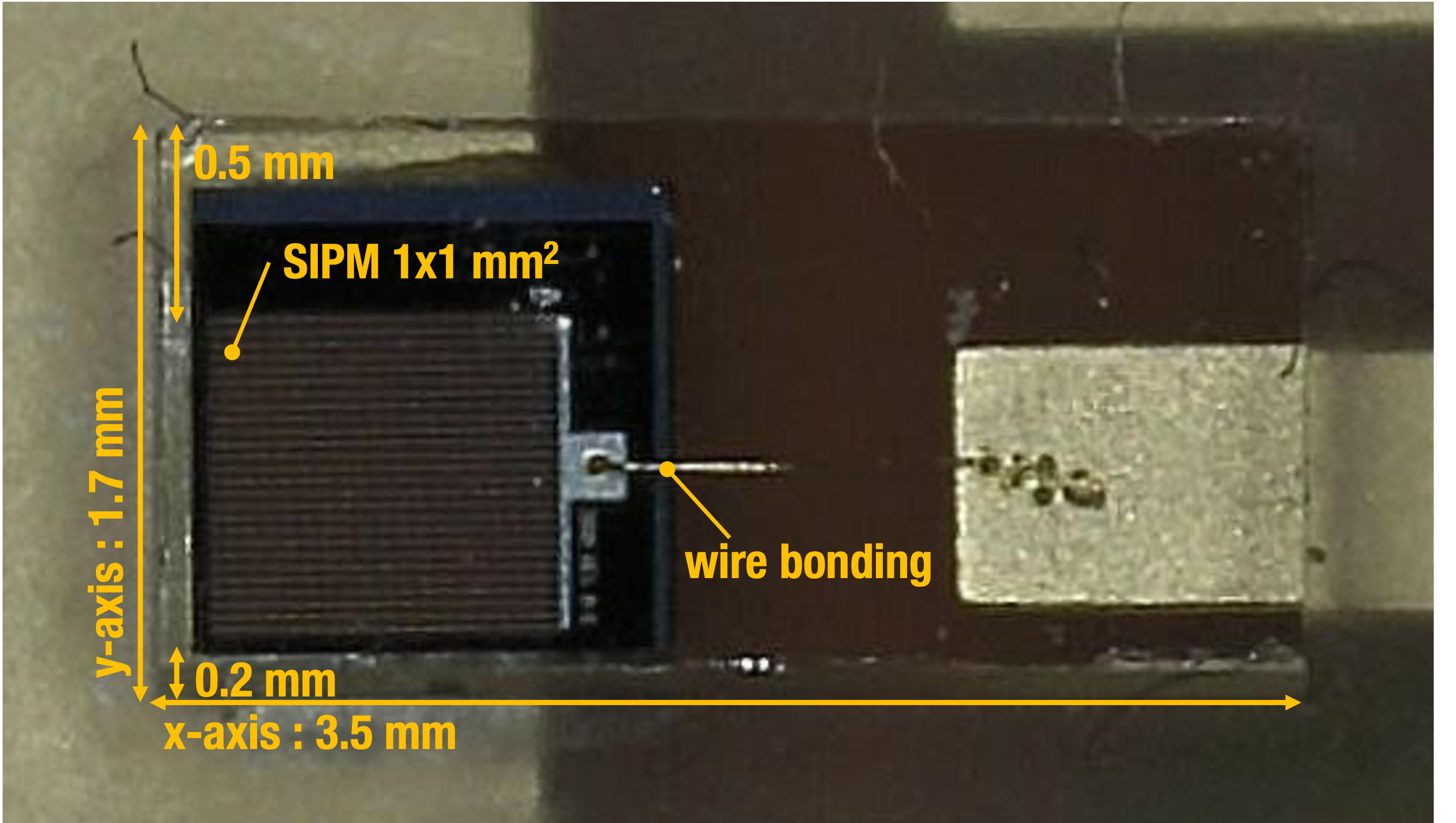}
        \caption{Photograph of the SiPM sensor with the area covered by the protection layer. Notice that in the test beam, the dimensions 1.7 mm $\times$ 3.5 mm correspond to the y and x coordinates, respectively.}
          \label{fig:FotoSiPM}
\end{figure}

Differently from \cite{SiPM1, SiPM2}, the prototypes with protection layers were no longer part of a six-sensor structure; instead, they were singly cut and the protection layer was applied with a more defined area as shown in Figure \ref{fig:FotoSiPM}. The area was 1.7 $\times$ 3.5 mm$^2$ , covering the wire bonding in the long horizontal direction (x-axis in Figure \ref{fig:FotoSiPM}), and  with the sensor not centered in the vertical short direction (y-axis in Figure \ref{fig:FotoSiPM}) but with -200 and +500 $\mu$m lateral  borders. This choice was dictated by the necessity to better control the effect observed in \cite{SiPM2} where particles also impinging outside the sensor surface could produce a signal due to the Cherenkov effect. For each protection layer, two equal sensors were measured in order to compare also possible different responses.

\subsection{Beam test setup} 
\label{subsec:tb}
The SiPMs response has been studied with MIPs (Minimum Ioinizing Particles) at the T10 beamline of CERN-PS in November 2022. The beam was mainly composed of protons and $\pi^+$. The momentum, where not stated differently, was of 10 GeV/c.

The telescope was made of four sensors: two SiPMs under test and two  LGAD detectors (1x1 mm$^2$ area and 35 $\mu m$ or 25 $\mu m$ thick prototypes) \cite{LGAD}; the latter were used as trigger for the beam particles and as timing reference. The four sensors were placed at 8 cm one from the other. Each detector was mounted on a remotely controlled moving structure capable of an x and y position precision of 10 $\mu$m: this configuration allowed  a precise study of the SiPMs response as a function of the position  of the beam. 

The whole setup was enclosed in a dark box at room temperature. The sensors were used without any cooling and the temperature ranged between 25 and 28 degrees.

The SiPMs signals were independently coupled to a customized front-end with an X-LEE amplifier\footnote{https://www.minicircuits.com/pdfs/LEE-39+.pdf}  featuring a gain factor of about 40 dB. The trigger was defined as the coincidence of the two LGADs in the telescope. At each trigger, all four waveforms were stored using a Lecroy Wave-Runner 9404M-MS digital oscilloscope.
%

The vertical scale of the oscilloscope was set to two different values: a larger one (200 mV/div) for the measurement of the number of fired SPADs, and a smaller one (100 mV/div) for timing measurements. 

\subsection{Signal selection}
\label{subsec:signal}

Signal and Dark Counts (DC) events selections proceed through different steps, similar to the procedure described in \cite{SiPM1, SiPM2}.

The SiPM signals are those within a window of $\pm$ 2 ns from the (t$_0$) trigger given by the coincidence of the two LGAD sensors. 
During the position scan (see section \ref{subsec:scanposition}) no further cuts were applied to evaluate the average signal amplitude. 

The DC signals are those measured in the interval -10 ns to -2.5 ns from the  t$_0$. It should be noticed that the SiPM performances and in particular the  Dark Count Rate (DCR) were surely affected by the sensors being operated at room temperature.
Moreover during the data taking an increase of the DCR with the integrated beam flux was observed: the effect was already noticed for the same type of detectors in \cite{Altamura, Acerbi}, but at energies of the protons around a few tens of MeV. In future, this effect should be quantified with dedicated measurements.



For the timing analysis, a fixed threshold of approximately 50$\%$ of the single SPAD signal amplitude was used for the tested SiPMs, while a CFD (Constant Fraction Discrimination) threshold of 50$\%$ was used for the reference LGADs.

\section{Results}\label{result}

\subsection{Position scan results}
\label{subsec:scanposition}
\begin{figure}
     \centering
     \begin{subfigure}[\label{fig:scan_x}]{0.6\textwidth}
         \includegraphics[width=\textwidth]{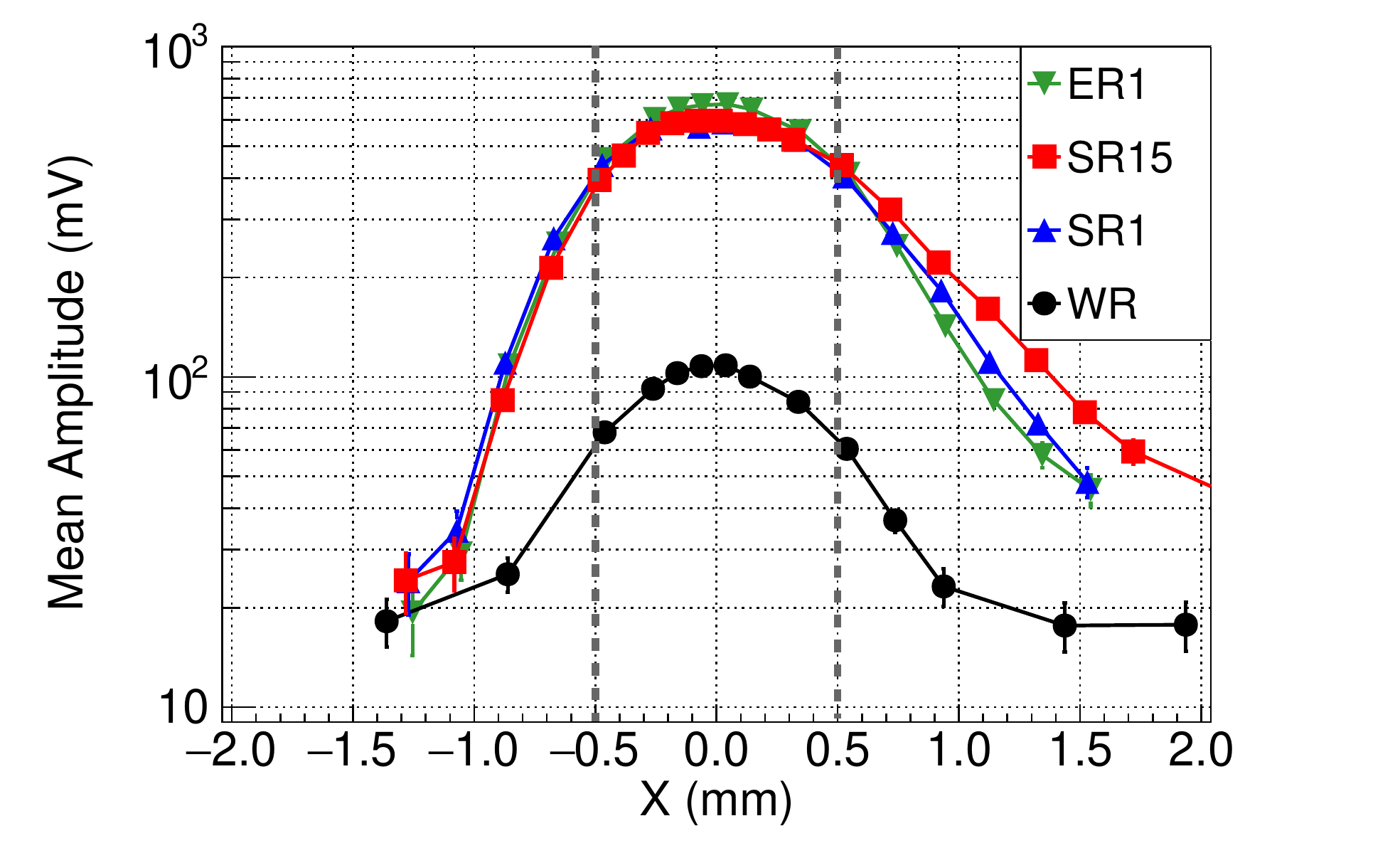}
     \end{subfigure}
     \begin{subfigure}[\label{fig:scan_y}]{0.6\textwidth}
         \includegraphics[width=\textwidth]{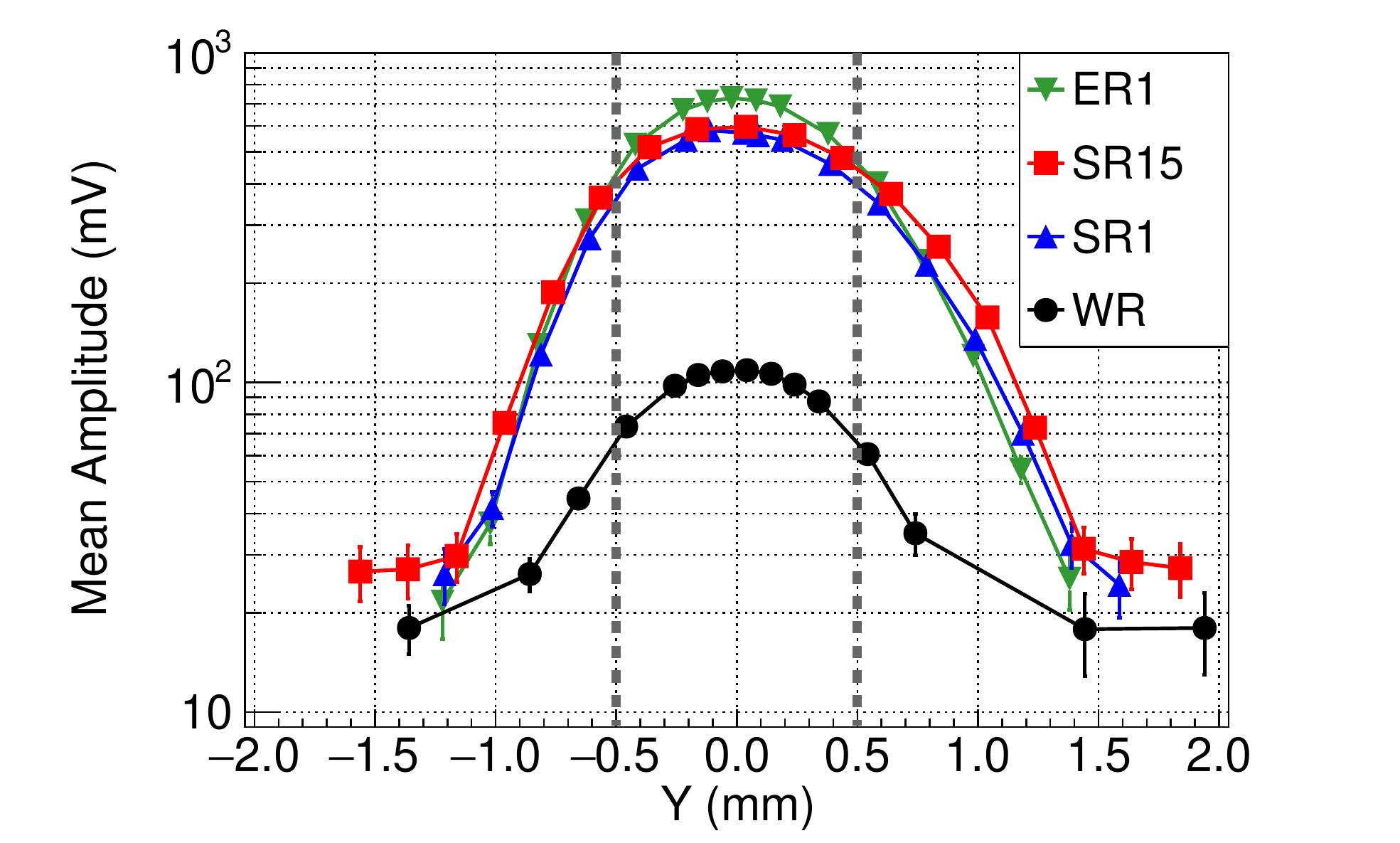}
     \end{subfigure}
\caption{Measured average signal as a function of the x (above) and y (below) with the position of the maximum setting the zero in the abscissa. The scans were performed at 3 V OV. The vertical dashed lines indicate the dimension of the SiPM.}
\label{fig:positionscan}
\end{figure}

In  Figure \ref{fig:positionscan} the average signal amplitude as a function of the different positions both for x and y scans are reported. Notice that the beam dimensions for the scan are defined by the LGAD coincidence trigger area, $\sim$1 mm$^2$, so it is expected to have not sharp edges, but convoluted with a smooth distribution as the scan proceeds. In each scan, the position of the maximum is referred to as the zero in the abscissa. The measurements were performed with the oscilloscope scale at 200 mV/div and at an Over-Voltage (OV) of 3 V, below the usual value in order to account for as much as possible fired SPADs.\\
Several observations might be done:
\begin{itemize}
    \item The average amplitude is much higher for the SiPM with protection layers with respect to the WR one, as expected due to the Cherenkov emission in the protection layer. 
    \item The WR sample is symmetric since only the sensor area is important.
    \item In the x and y scan for SR and ER sensors, the distributions are larger than WR because of the Cherenkov light from particles outside the sensor area but inside the protection layers. Moreover, they are asymmetric due to the different borders of the protection layers with respect to the sensor (see section \ref{sec:detectors}). 
    \item Notice that outside the SiPM area the red curves, corresponding to SR15, are systematically higher with respect to the other due to the thicker layer: at the maximum this cannot be appreciated because, although the cone radius is different, the limit of detection is given by the SiPM area. 
    \item On the tails, all distributions are reduced to the simple DCR value. However, due to the extension of the beam (FWHM about 1 cm), the illuminated area is bigger than the triggered one and a small residual signal can be observed in the SR or ER samples.
    \item No measurable difference between the scan of two different sensors with the same protection layer material and thickness was observed, underlying the repeatability of the measurements. 
\end{itemize}

\begin{figure}[h!]
        \centering%
        \includegraphics [width=1\textwidth]{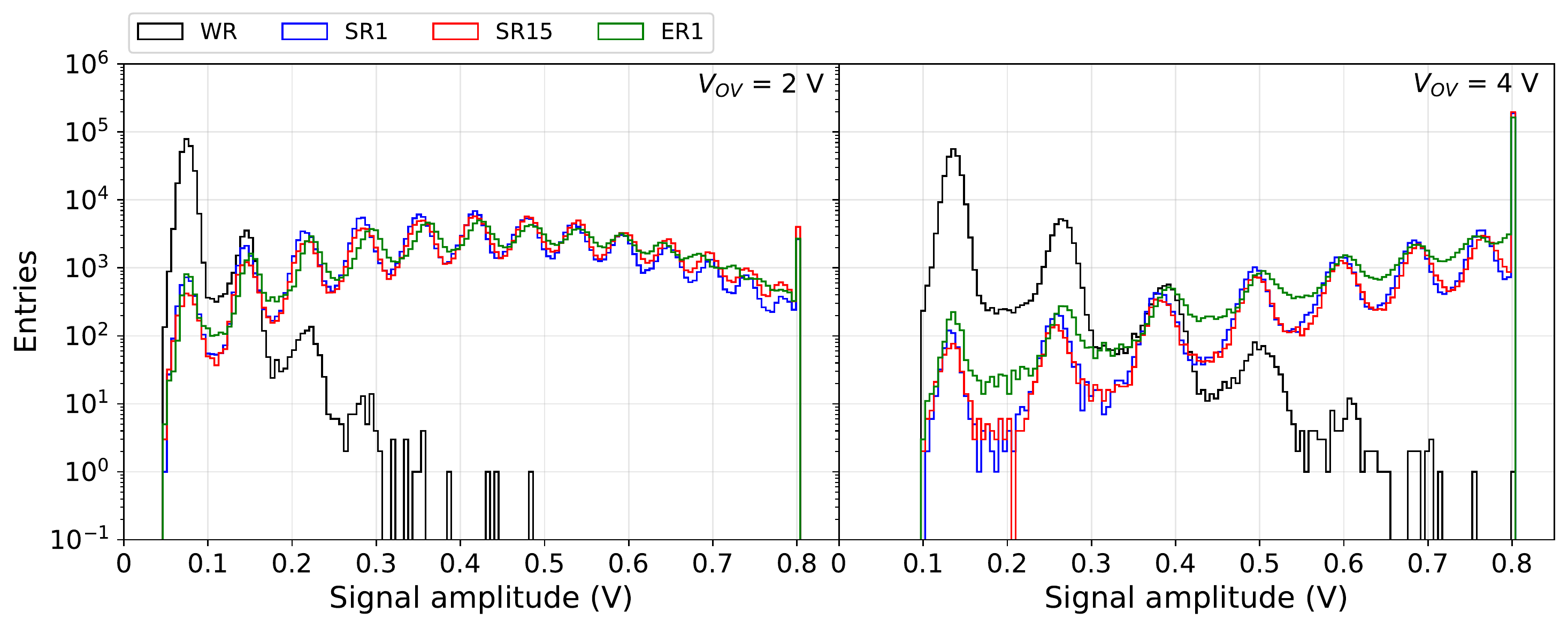}
        \caption{Distribution of the amplitude for different sensors as measured at the position of the maximum signal average at OV (left) 2 V and (right) 4 V. The curves are obtained with an oscilloscope vertical scale at 100 mV/div. The plots are normalized to the same number of events.} 
          \label{fig:distributionatmax}
\end{figure}

In  Figure \ref{fig:distributionatmax} the distributions of the signals amplitudes are compared as obtained at the position of the maximum in Figure \ref{fig:positionscan}. The results refer to 2 V and 4 V of OV (both 
with oscilloscope vertical scale at 100 mV/div) and different protection layers, with curves normalized to the same number of events and an amplitude scale, re-normalized at the first peak. For this plot, a baseline reference was evaluated and subtracted from the signals. The baseline  was determined as the average amplitude in the interval -10 ns to -2.5 ns from the  t$_0$. Since a wrong evaluation of the baseline could bring to a fake and unwanted enlargement of the peaks, to allow a clearer comparison, events in this control region with amplitudes   $>$10$\%$ of the single SPAD were rejected since contaminated by DC signals. 

In  Figure \ref{fig:distributionatmax} it can be observed that the number of peaks present in the curves (corresponding to the number of SPADs fired) is much higher in the sensors with protection than in the WR sensor, one more time underlying the huge contribution of the Cherenkov light.
Notice that for the WR sensor the peaks beyond the first SPAD are compatible with the expected CT related to a single SPAD firing (see also \cite{SiPM2}). It should be underlined that at higher OV the peaks of the distributions move to higher values, even enhancing the number of overflows, due to the increase of the gain. The same figure, at the lower OV where the full spectrum is more visible, indicates that the SR15 is slightly higher than the SR1, as expected from the thicker protection layer. There is no marked difference between ER and SR sensors. 

These results confirm the importance of a protection layer to enhance the response of SiPM sensors to a traversing charged particle.  The different thickness of the layer produces a different cone radius for the photons emission and an increased number of photons as indeed is observed in Figure \ref{fig:positionscan},\ref{fig:distributionatmax}. The ER/SR comparison is not conclusive with the present data. However, the effective evaluation of the number of firing SPADs should be better studied using a larger SiPM area or an array of SiPMs in order to cover the entire area of the photons produced by Cherenkov emission. 

\subsection{Results at lower energies}
\label{subsec:lowenergy}

It is interesting to try to compare proton response as a function of the momentum; indeed, it is expected that the light produced by the charged particle  passing  through the protective layer should be reduced when approaching the Cherenkov threshold (around 0.7 GeV/c for protons). 

The T10 beam at CERN-PS has the possibility to provide particles down to 1 GeV/c although the nominal species composition strongly varies: it is dominated by protons and pions (70/20\%)  at high energies, while a considerable fraction of positron is present at  1 GeV/c \cite{vandijk}.
The momentum resolution is estimated to be $\sim$5\%. 

To select protons also at such a low energy the excellent timing performances of the LGAD trigger sensors (24 cm apart) were exploited.

\begin{figure}[hbt]
     \centering
     \begin{subfigure}[]{0.5\textwidth}
         \includegraphics[width=\textwidth]{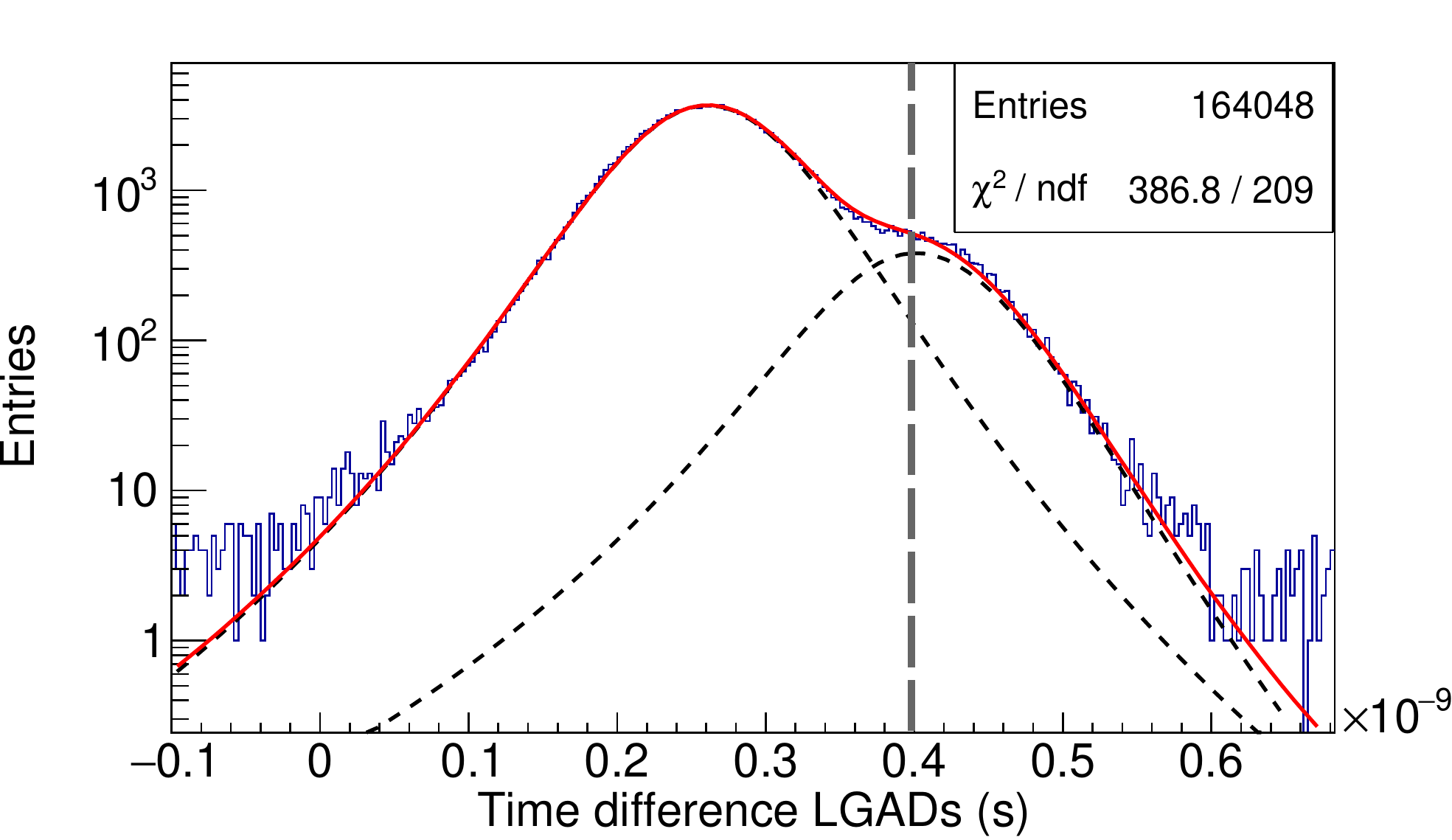}
         \caption{}
    \label{fig:tof_15}
     \end{subfigure}
     \hspace{-0.2cm}
\begin{subfigure}[]{0.5\textwidth}
         \includegraphics[width=\textwidth]{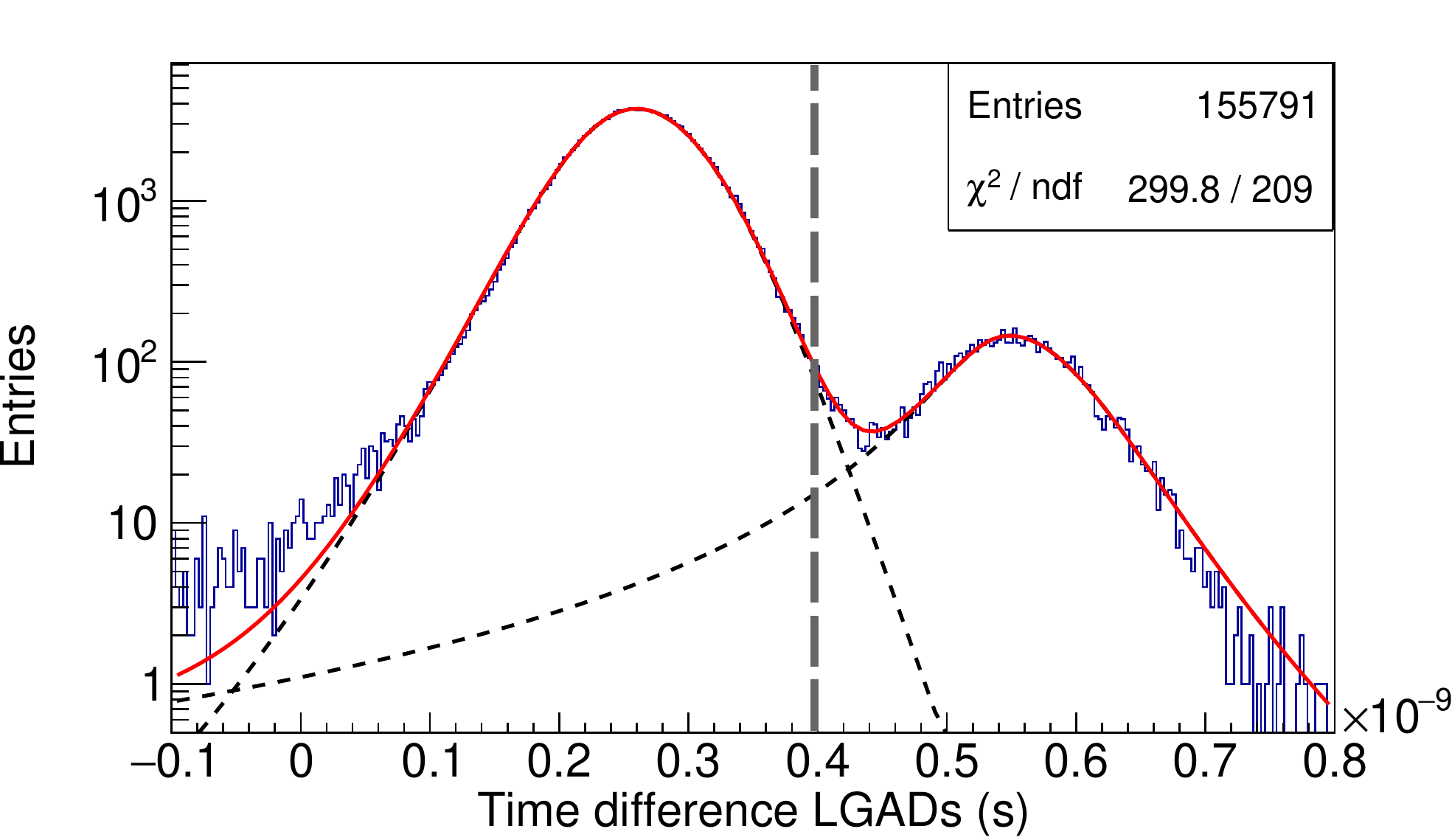}
        \caption{}
       \label{fig:tof_1}
     \end{subfigure}
\caption{Measured time differences between first and last LGAD sensor for 1.5 GeV/c (a) and at 1.0 GeV/c (b). The fit is performed with the sum of two q-Gaussians. The dashed vertical line at 0.4 $10^{-9}$s indicates the cut applied in order to select protons (more details in the text).}
\end{figure}

In Figure \ref{fig:tof_15} and \ref{fig:tof_1} the difference of the measured time between the two LGADs is reported  for 1.5 and 1.0 GeV/c respectively. The curve is fitted considering the sum of two q-Gaussian distributions. The peak on the right represents protons while the peak on the left corresponds to electrons and pions. The cut to discriminate electrons and pions from protons is made considering a distance of 3$\sigma$ from the mean of the electrons/pions peak. Choosing events with a time difference $\geq$ 0.40 ns ( 1 GeV/c)  and between 0.40 and 0.68 ns (1.5 GeV/c), it is possible to select protons with an estimated purity of 82\% and 84\% respectively. In total about 4000 and 6500 events were selected at 1.0 and 1.5 GeV/c respectively.

\begin{figure}[h!]
        \centering%
        \includegraphics [width=0.8\textwidth]{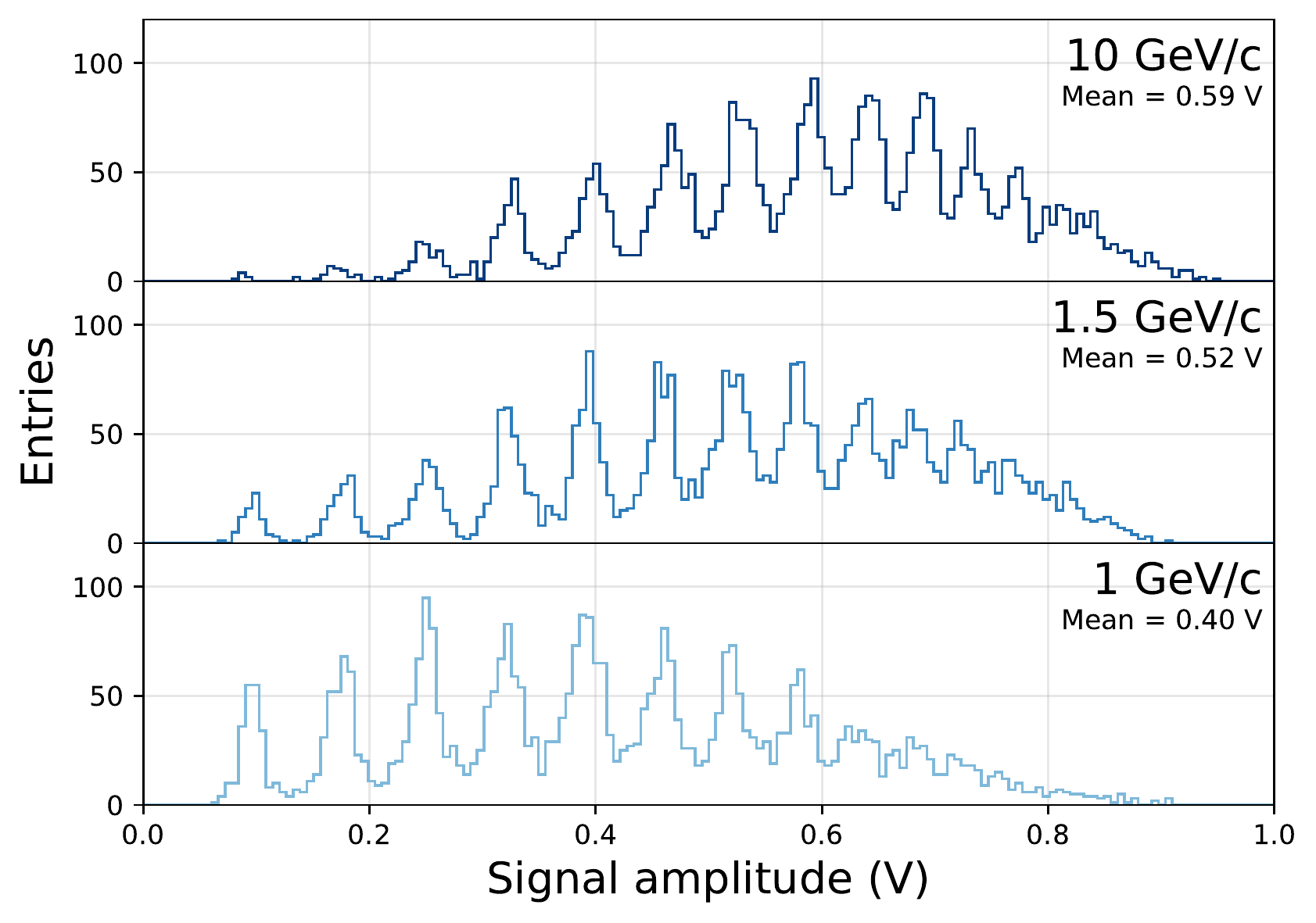}
        \caption{Distribution of the detected number of SPADs for sensor SR1 at 10.0, 1.5 and 1.0 GeV/c. The plots are obtained at 3 V OV and with an oscilloscope vertical scale at 200 mV/div.}
          \label{fig:scanlowAMP}
\end{figure}

Figure \ref{fig:scanlowAMP} reports 
the number of fired SPADs distribution for the SiPM sensor SR1 at three different proton energies: 10.0, 1.5  and 1.0 GeV/c. The same number of events were used for the three plots. A baseline subtraction, similar to Figure \ref{fig:distributionatmax}, was applied.

It is expected that the amount of Cherenkov radiation emitted depends on the particle momentum, and indeed in the figure it is possible to observe a reduction in the detected average signal as the beam approaches the proton Cherenkov threshold. Still, at 1 GeV/c a mean number of firing SPADs of $\simeq 5$ is measured.

\subsection{Timing results}
\label{subsec:timing}

\begin{figure}[h!!]
        \centering
        \includegraphics[width=0.8\textwidth]{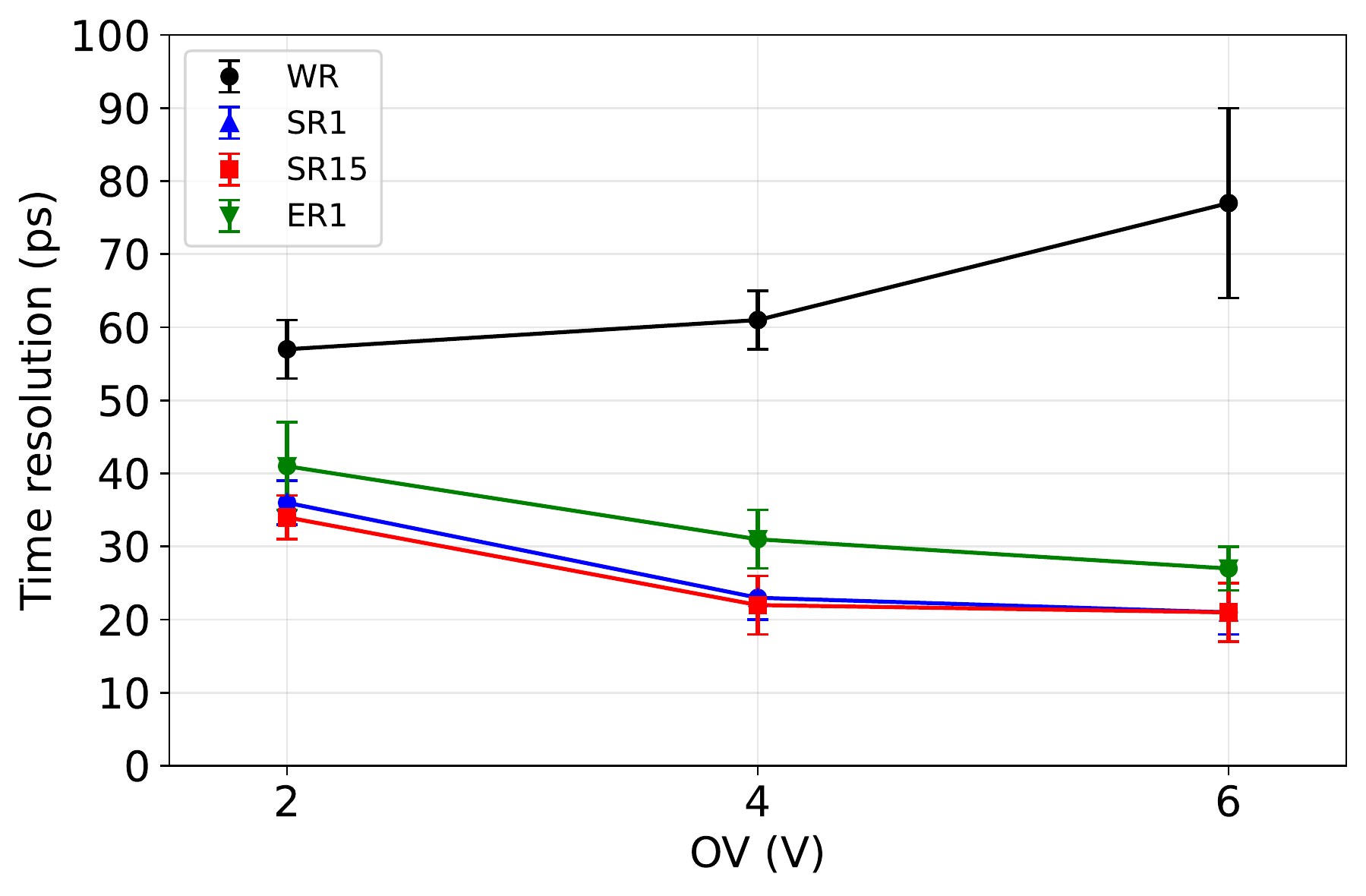}
        \caption{Measured time resolution as a function of the over-voltage OV for the sensors  described in Section \ref{sec2}. 
        }
          \label{fig:timing}
\end{figure}

\begin{figure}[h!!]
        \centering
        \includegraphics[width=0.8\textwidth]{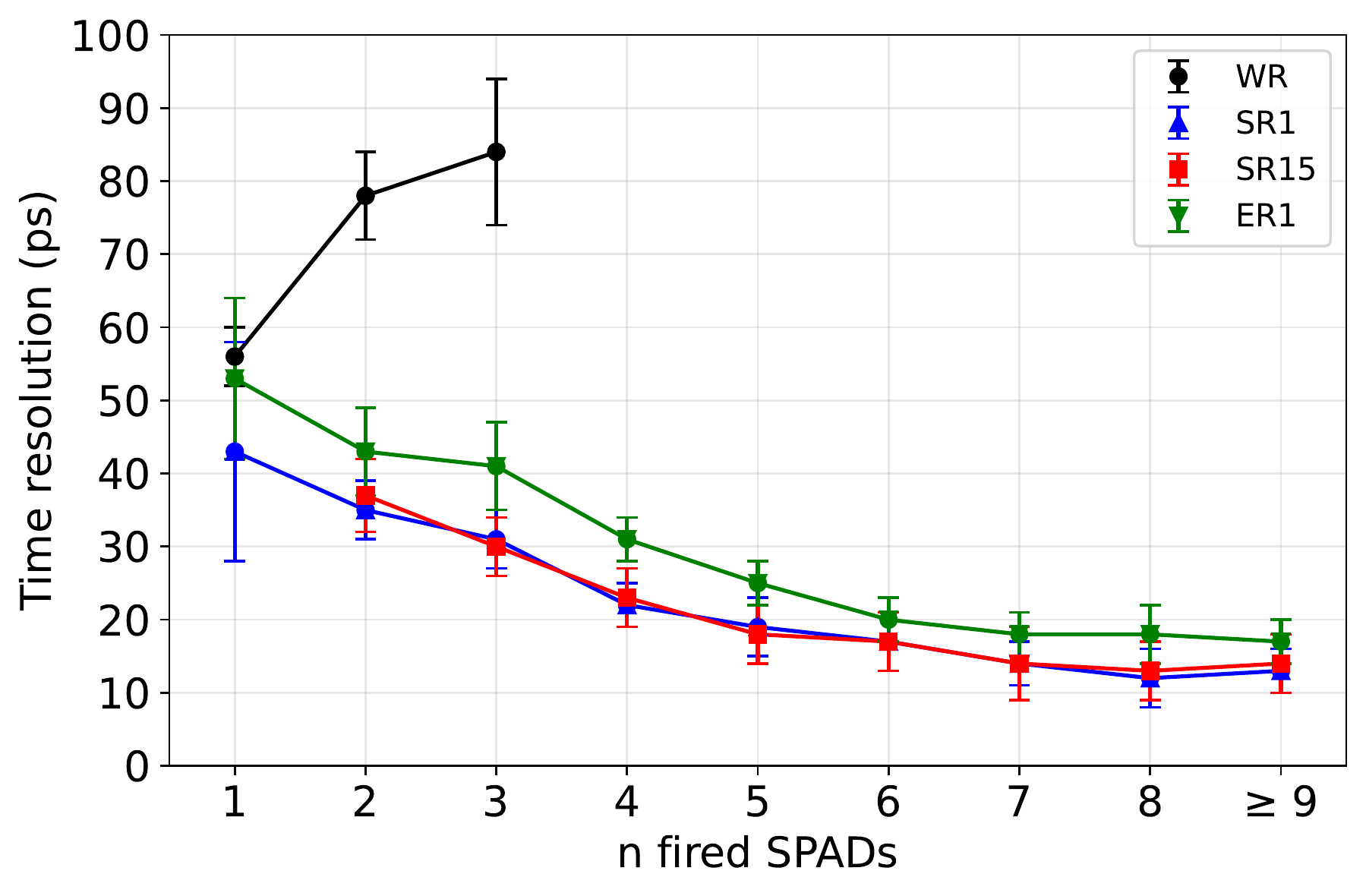}
        \caption{Measured time resolution as a function of the number n of firing SPADs for the sensors described in Section \ref{sec2} at 4 V OV.  The last value of WR is intended to be for signals with $\geq$3 SPADs firing.}
          \label{fig:timingSPAD}
\end{figure}

The measured time resolution at 10 GeV/c is reported for the different sensors tested as a function of the over-voltage (Figure \ref{fig:timing}) and  observed number of SPADs, at an over-voltage of $\sim$ 4 V (Figure \ref{fig:timingSPAD}). The measurements were done with data collected at the maximum of the position scan distribution, i.e. with the SiPM aligned with the center of the beam. The results were obtained using one of the LGADs as reference, measuring the time difference sensor-LGAD, evaluating the sigma from a q-Gaussian fit  and subtracting the LGAD contribution, typically of 31 ps as in \cite{SiPM1}. The threshold of the signal from the SiPM sensor was set to 50\% of a single SPAD amplitude. Baseline subtraction was applied as for Figure \ref{fig:distributionatmax}. Relaxing these selection criteria worsens the results since some DC noise is introduced: the WR results get worst by around 15 ps, while for the sensors with resins, the time resolution increases by 3 to 5 ps.  The quoted errors include the statistics and an evaluation of the systematic uncertainties estimated via a variation of the signal threshold and a variation on the reference LGAD timing. All measurements are the average of the two different samples available for each sensor.

Figure \ref{fig:timing} shows that the three sensors with protection layers behave in a similar way within the errors, reaching values around 20-30 ps at the maximum overvoltage, a value slightly better than \cite{SiPM2} due to the better alignment obtained. The WR sensors show an increase in time resolution with OV probably related to the increase of DCR with OV (see below). 

  
\begin{table}
\centering
\caption {Time resolution with respect to a selected number n of firing SPADs of different types of SiPMs at 4 V OV. The percentage $F_n$ is the rounded mean fraction of events with a signal corresponding to $n$ SPADs with respect to the total. The last value of WR is intended to be for signals with $\geq$3 SPADs firing. }
\label{tab:trvalues4}
\begin {tabular}{ccccccccc}
\toprule
       & \multicolumn{2}{c}{SR15} & \multicolumn{2}{c}{SR1}  & \multicolumn{2}{c}{ER1}  & \multicolumn{2}{c}{WR}  \\
 \midrule
n & $\sigma$ (ps) & $F_n$($\%$)  &$\sigma$ (ps) & $F_n$($\%$)  &$\sigma$ (ps) & $F_n$($\%$) & $\sigma$ (ps) & $F_n$($\%$) \\ 
SPADs & & & & & & & & \\  \hline
1 &  - & $\sim$0 &  $43 \pm 15 $  & $\sim$0 & $53 \pm 11$ & $\lesssim$1 &  $56 \pm 4$ & $\sim$81 \\
2 &  $37 \pm 5$ &  $\sim $0 &  $35 \pm 4 $  &  $\lesssim $1 & $43 \pm 6$ &  $\sim$1 &  $78 \pm 6$ & $\sim$16  \\
3 &  $30 \pm 4$ & $\lesssim $1  &   $31 \pm 4  $  & $\sim$1 & $41 \pm 6$  & $\lesssim$3  &  $84 \pm 10$ & $\sim$3 \\
4 &  $23 \pm 4$   & $\lesssim$2 &  $22 \pm 3 $  & $\lesssim$3  & $31 \pm 3$ & $\sim$5  & - &  - \\
5 &  $18 \pm 4$  & $\lesssim$4  &   $19 \pm 4 $   & $\lesssim$5 & $25 \pm 3$  & $\sim$8  & - & - \\
6 &  $17 \pm 4$  & $\sim$5 &  $17 \pm 4 $ & $\lesssim$7  & $20 \pm 3$ & $\lesssim$12  & - &  - \\
7 &  $14 \pm 5$  & $\lesssim$8 &   $14 \pm 3 $ & $\lesssim$10  & $18 \pm 3$ & $\lesssim$13  &  - & - \\
8 &  $13 \pm 4$   & $\sim$9 &  $12 \pm 4 $  & $\sim$11 & $18 \pm 4$ & $\sim$10 & - &  - \\
$\geq $9 &  $14 \pm 4$   & $\sim$71 &  $13 \pm 3 $ & $\sim$63 &  $17 \pm 3$ & $\sim$48 &  - & -\\
\bottomrule

\end{tabular}
\end{table}

As reported in Figure \ref{fig:timingSPAD} and Table \ref{tab:trvalues4}, in the present work it was possible to extend the  measurement of \cite{SiPM1} and \cite{SiPM2}  to a higher number of SPADs, confirming the improvement of the time resolution with the number of SPADs fired, compatible with a naive scaling $\frac{1}{\sqrt{n_{\text{SPAD}}}}$. Notice that in the analysis of time resolution versus number of fired SPADs, each point is individually analyzed, employing ad hoc calibration and correction..

Time resolutions better than 20 ps are obtained for n SPADs $\geq$ 6. The results indicate that the number of SPADs is the main contributor to the improvement of the time resolution. Moreover, for the sensors with protection layers, the fraction of events with a small number of SPADs is low (see Table \ref{tab:trvalues4}). As a consequence, such detectors would have a huge noise rejection, when compared to the standard use of SiPMs in detecting photons: indeed a threshold above 3 SPADs would keep an efficiency higher than 99\% making rid of the noise due to DCR and correlated CT.
Note that the limitations of the present Front End electronic jitter contribution to the timing resolution may imply a limit in the values obtained at high number of SPADs per event. Similar results are obtained at 6 OV. For the WR sensor, results at 2 and 3 fired SPADs are strongly affected by the DCR and CT  that deteriorate the measurement.


\section{Conclusions}\label{conclusion}

In this paper, the response of SiPMs to the passage of a charged particle  was further studied for several sensors produced by FBK, with and without different protection layers. By means of an accurate position scan it was possible to study and characterize the response of the sensors. 

The measurements confirm the strong production of photons due to the Cherenkov effect in the standard sensor protective layer. In addition to this, it was possible to appreciate differences related to the thickness: a larger production of photons for the thicker layer is observed related to the larger cone radius for photon production. The epoxy and silicon-based resins do not show a difference in the number of photons produced. Notice that, in presence of this effect, the SiPM fill factor is negligible for the efficiency.

The effect of reduced photon yields near the Cherenkov threshold for protons was observed varying  the beam energy from 10 to 1 GeV/c. 

The timing performances were measured, indicating a value of about 20-30 ps at 4-6 V of OV for MIPs at 10 GeV/c. The improvement as a function of the number of SPADs fired is confirmed, reaching values below 20 ps for six or more fired SPADs (the majority of cases). 

In future, it would be important to repeat the measurements with larger matrices in order to exploit all possible photons produced. Moreover, the effect of radiation damage should be better quantified, especially in view of possible new applications in hostile environments, like combined TOF RICH systems (as in ALICE 3) or TOF systems for space experiments. However, these results pave the way for the use of SiPM as sensors for the direct detection of charged particles with high precision timing.


\section*{Declarations}
The study was funded by: INFN and FBK.
The authors received research support from institutes as specified in the author list beneath the title. \\

\section*{Data availability}
The datasets generated during and/or analysed during the current study are available from the corresponding author on reasonable request.

\bibliography{sn-bibliography}

\begin{thebibliography}{1}
\providecommand{\url}[1]{{#1}}
\providecommand{\urlprefix}{URL }
\providecommand{\doi}[1]{\url{https://doi.org/#1}}
\bibcommenthead

\bibitem{SiPM1}
F.~Carnesecchi, et~al., Direct detection of charged particles with {SiPMs}.
\newblock Journal of Instrumentation. \textbf{17}(P06007) (2022).
\newblock \doi{10.1088/1748-0221/17/06/P06007}

\bibitem{SiPM2}
F.~Carnesecchi, et~al., Understanding the direct detection of charged particles
  with {SiPMs}.
\newblock Eur. Phys. J. Plus \textbf{138}, 337 (2023).
\newblock \doi{10.1140/epjp/s13360-023-03923-4}

\bibitem{ALICE3}
{Letter of intent for ALICE 3: A next generation heavy-ion experiment at the
  LHC} (2022).
\newblock \doi{10.48550/arXiv.2211.02491}

\bibitem{2020Mazzi}
A.~Mazzi, et~al.
\newblock {SiPM development at FBK for the barrel timing layer of CMS} (2020).
\newblock \url{https://indico.cern.ch/event/813597/contributions/3727862/}

\bibitem{2019Gola}
A.~Gola, et~al., {NUV-Sensitive Silicon Photomultiplier Technologies Developed
  at Fondazione Bruno Kessler}.
\newblock {Sensors} \textbf{19} (2019).
\newblock \doi{10.3390/s19020308}

\bibitem{LGAD}
F.~Carnesecchi, et~al., Beam test results of 25 $\mu$m and 35 $\mu$m thick
  {FBK} ultra fast silicon detectors.
\newblock The European Physical Journal Plus \textbf{138}(99) (2023).
\newblock \doi{10.1140/epjp/s13360-022-03619-1}

\bibitem{Altamura}
A.~Altamura, et~al., Characterization of {S}ilicon {P}hotomultipliers after
  proton irradiation up to 10$^{12}$neq{/cm$^2$}.
\newblock Nuclear Instruments and Methods in Physics Research Section A
  \textbf{1040} (October 2022).
\newblock \doi{10.1016/j.nima.2022.167284}

\bibitem{Acerbi}
F.~Acerbi, et~al., Radiation damage effects of protons and x-rays on silicon
  photomultipliers.
\newblock Nuclear Instruments and Methods in Physics Research Section A
  \textbf{1047} (February 2023).
\newblock \doi{10.1016/j.nima.2022.167791}

\bibitem{vandijk}
M.~Van~Dijk, et~al.
\newblock {Introduction to Secondary Beams} (2023).
\newblock \url{https://indico.cern.ch/event/1254858/}

\end{thebibliography}


\end{document}